\documentclass[aps,showpacs,twocolumn]{revtex4}
\usepackage{amsfonts}
\usepackage{amsmath}
\usepackage{amssymb}
\usepackage{graphicx}

\begin{document}

\title{Emergence of Topological Fermi Liquid from a Strongly Correlated
Bosonic System in Optical Superlattices}
\author{Bo-lun Chen}
\affiliation{Department of Physics, Beijing Normal University, Beijing 100875, China}
\author{Su-peng Kou}
\thanks{Corresponding author}
\email{spkou@bnu.edu.cn}
\affiliation{Department of Physics, Beijing Normal University, Beijing 100875, China}
\keywords{}
\pacs{03.75.Hh, 03.75.Lm, 64.60.Cn, 74.20.-z, 74.20.Mn}

\begin{abstract}
Recent experiments on quantum degenerate gases give an opportunity for
simulating strongly-correlated electronic systems in optical lattices. It
may shed light on some long-standing puzzles in condensed-matter physics,
like the nature of high-temperature superconductivity in cuprates that had
baffled people over two decades. It is believed that the two-dimensional
fermionic Hubbard model, or $t$-$J$ model, contains the key to this problem;
but the difficulty of unveiling the mystery of a strongly-interacting
fermionic system is also generally acknowledged. Here, as a substitute, we
systematically analyze the property of bosonic $t$-$J$ model simulated in
optical superlattices near unit-filling. In particular, we show the
emergence of a strange topological Fermi liquid with Fermi surfaces from a
purely bosonic system. We also discuss the possibility of observing these
phenomena in ultracold atom experiments. The result may provide some\
crucial insights into the origin of high-$T_{c}$ superconductivity.
\end{abstract}

\maketitle

In recent years, the physics community has witnessed a series of exciting
discoveries and achievements. Among them, using ultracold atoms that form
Bose-Einstein Condensates (BEC) or Fermi degenerate gases to make precise
measurements and\ simulations of quantum many-body systems, is quite
impressive and has become a rapidly-developing field\cite{Bloch's
review,review}. Since atoms are cooled down to temperature\ near\ absolute
zero and trapped in optical lattices building from six orthogonal laser
beams, they provide us a peaceful playground for manipulating atoms with
unprecedented accuracy. Some pioneering works\cite{Jaksch} revealed the
promising potential of applying ultracold atoms to make quantum computer and
quantum simulator: By changing the intensity, phase and polarization of
incident laser beams, one can tune the Hamiltonian parameters including the
dimension, the hopping strength and the particle interaction at will. People
have successfully observed the Mott insulator--superfluid transition in both
bosonic\cite{Greiner} and fermionic\cite{Esslinger,Bloch2} degenerate gases,
and have demonstrated how to produce\cite{Duan} and control effective spin
interactions in a double-well ensemble\cite{Bloch}. All these evidences
imply that an era in which atomic and optical physics unites with
condensed-matter physics is within sight.

Particularly, the two-dimensional fermionic Hubbard model (or $t$-$J$ model)
is one of the most interesting issues depicting the nature of
high-temperature superconductivity\cite{hir,anderson,ZR}. Ever since the
discovery of high $T_{c}$ cuprates, tremendous efforts had been contributed
to investigations of this model. Derived from Hubbard model at half-filling
(one electron per site), the $t$-$J$ model describes the motion of doped
holes in an antiferromagnetic (AF) background. It carries the essence of a
strongly correlated electronic system with intrinsic competition between
superexchange interaction of spins and hopping processes of charge-carriers
(holes). Over the past two decades, people have employed many methods and
developed different schemes, hoping to fully understand this model. Although
a lot of consensus have been accomplished, there are still some ambiguities
to be clarified. For example, can superconductivity evolve from a purely
fermionic repulsive many-body system? Are there exist some new phases of
matter in high-$T_{c}$ superconductors?

Therefore, some people suggest to use its bosonic counterpart, the bosonic $%
t $-$J$ model\cite{Boninsegni1,Boninsegni2,mean field phase}, as a trial
model to investigate, for bosons are much more easier to deal with in both
analytic and numerical approaches. After obtaining some experiences and
conclusions, we may use them as\ reminders and analogies to original
fermionic model. Besides, considering the present situation of experiments
in ultracold atoms, where bosons are more accessible to be cooled and
controlled, we believe it is worthwhile to explore the bosonic $t$-$J$ model
in optical lattices which can be formally written into two parts:
\begin{equation}
\hat{H}=\hat{H}_{t}+\hat{H}_{J}=-t\sum_{\langle ij\rangle \sigma }(\hat{a}%
_{i\sigma }^{\dagger }\hat{a}_{j\sigma }+H.c.)+J\sum_{\langle ij\rangle }%
\mathbf{\hat{S}}_{i}\cdot \mathbf{\hat{S}}_{j},  \label{bosonic tJ}
\end{equation}%
where Hilbert space is restricted by the no-double-occupancy constraint $%
\sum_{\sigma }\hat{a}_{i\sigma }^{\dagger }\hat{a}_{i\sigma }\leq 1$, $\hat{a%
}_{i\sigma }$ annihilates a two-component boson ($\sigma \equiv \uparrow
,\downarrow $ denoting two internal states) and $\mathbf{\hat{S}}_{i}$ is
the (pseudo)spin operator at site $i$, $\mathbf{\hat{S}}_{i}\boldsymbol{=}%
\frac{1}{2}\sum_{\alpha \beta }\hat{a}_{i\alpha }^{\dag }\mathbf{\sigma }%
_{\alpha \beta }\hat{a}_{i\beta }$ with Pauli matrix $\mathbf{\sigma }%
=\left( \sigma ^{x},\sigma ^{y},\sigma ^{z}\right) $, $\langle ij\rangle $
denotes the nearest-neighboring counting. $\hat{H}_{t}$ and $\hat{H}_{J}$
describes hopping ($t>0$) and an effective AF superexchange ($J>0$)
interactions respectively.

In a seminal experiment\cite{Bloch}, Bloch \textit{et al.} loaded a
two-component $^{87}$Rb condensate, $\left\vert F=1,m_{F}=1\right\rangle
(\uparrow )$ and $\left\vert F=1,m_{F}=-1\right\rangle (\downarrow )$, into
arrays of isolated double wells. By changing the relative phase of the two
laser standing\ waves, the potential difference $\Delta $ in one double-well
can be raised or ramped down. The oscillation of the condensate after such
manipulation contained information of the strength and sign of the effective
superexchange energy $J$. With proper bias $\Delta $, they could convert
ferromagnetic interaction ($J<0$) into antiferromagnetic one ($J>0$).

\begin{figure}[tbp]
\begin{center}
\includegraphics[width=0.5\textwidth]{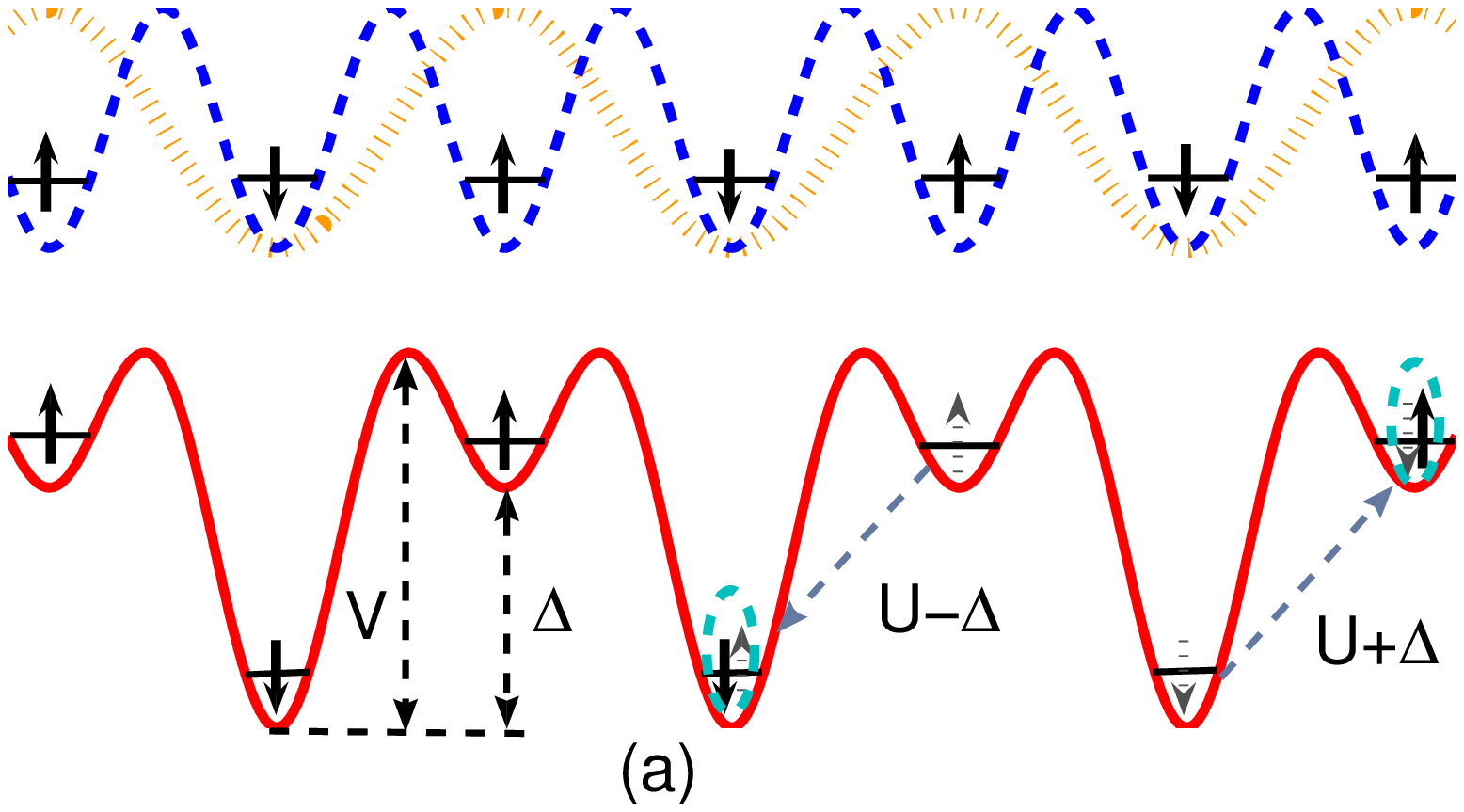} %
\includegraphics[width=0.5\textwidth]{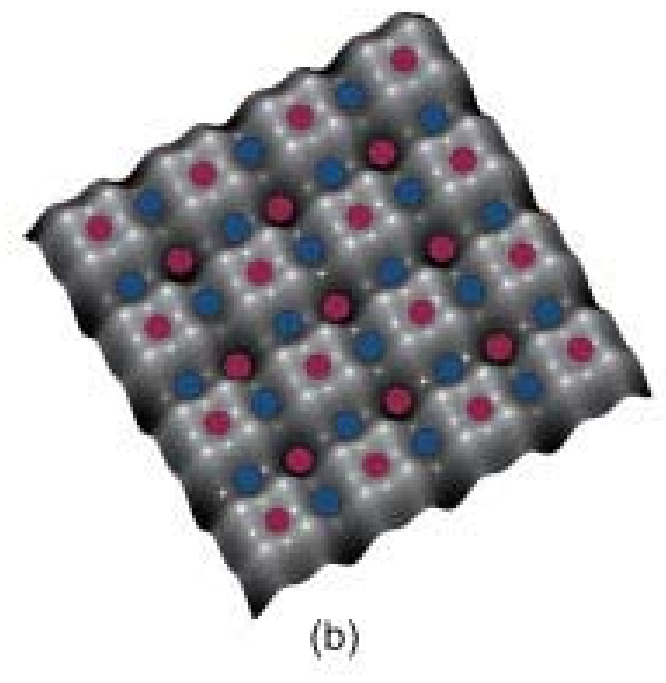}
\end{center}
\caption{Illustrations of optical superlattices. (a) Construction of an
optical superlattice. Combining two sets of optical lattices (wavelength $%
\protect\lambda $ and $2\protect\lambda $) with a relative phase difference $%
\protect\pi $ to produce a superlattice. $V$ is the depth of the potential
well. The superexchange processes are realized through virtual hoppings,
resulting perturbations $U\pm \Delta $ in the Hamiltonian. (b) A
visualization of a 2D superlattice, blue atoms denote one spin species ($%
\left\vert \uparrow \right\rangle $) and red atoms denote the other species (%
$\left\vert \downarrow \right\rangle $).}
\label{superlattice}
\end{figure}

To implement a controllable hopping term, we have to generalize the
double-wells to a set of biased optical superlattices. As illustrated in
Fig. (\ref{superlattice}), for instance, we can start from the situation
with one boson per site, different species in neighboring sites. Then by
adding another set of optical lattice adiabatically, whose wavelength is as
twice as the original one and has a relative phase difference of $\pi $, we
can build a superlattice (with shallow sites and deep sites), where a
certain type of bosons being trapped in a certain type of sites, for
instance, $\left\vert \uparrow \right\rangle $ in shallow sites and $%
\left\vert \downarrow \right\rangle $ in its neighboring deep sites.

By changing the shape of the superlattice, one can lift the potential
difference $\Delta $ to a fixed value to realize the effective AF
superexchange interaction ($J>0$). Furthermore, by adjusting the relative
phase and polarization between incident lasers, one can control the
tunneling amplitude $t_{i\sigma }$ of different inner states of atoms in
different sites to be identical. Therefore, we can introduce the hopping
effect while keeping a(n) (inhomogeneous) Heisenberg-type interaction in an
optical superlattice. (see Methods for details.)

In addition, according to ref. \cite{Ott}, one can introduce vacancies in a
BEC trapped in one-dimensional (1D) optical lattice by pointing an electron
beam at specific sites to remove atoms. Due to the small diameter ($100\sim
150$ nm) of the electron beam compared with the lattice spacing, holes can
be doped in without additional losses in other sites. The final occupying
configuration can be visualized with highly spatial resolution through
scanning electron microscopy. Since single-occupancy has been achieved in
current experiments\cite{Bloch,single occupancy} as an initial setup, we
believe that this technique can be further extended to 2D and 3D
superlattices loaded with multi-component atoms, thus one can manipulate
and(or) remove different species deliberately and observe the dynamics of
holes with single-atom, single-site sensitivity. Above all, the realization
of theoretical model (\ref{bosonic tJ}) as well as the necessary doping
technique are all within reach in current experiments.

The main analysis is arranged as follows: for bosonic $t$-$J$ Hamiltonian
near unit-filling, we analyze the doping effect in a topological
perspective. We show the existence of several exotic phases as vacancies
being gradually introduced. Then we discuss how to detect these features in
currently-available techniques in ultracold atom experiments.

\section{Holons: bosons or fermions?}

At unit-filling, the bosonic $t$-$J$ model is simple and can be reduced to
AF Heisenberg model, $\hat{H}=J\sum_{\langle ij\rangle }\mathbf{\hat{S}}%
_{i}\cdot \mathbf{\hat{S}}_{j}$. A variational wave-function based on
Resonating Valence Bond (RVB) picture by using Schwinger-boson description
can produce\cite{liang,chen,aa} an unrivaled accurate ground-state energy; a
generalized version\cite{chen} can further provide the staggered
magnetization and spin excitation spectrum precisely. So the bosonic RVB
picture is a natural choice for describing AF Heisenberg model. Explicitly,
we can introduced two flavors of bosons on each site, created by a canonical
operator $\hat{b}_{i\sigma }^{\dagger }$ acting on the vacuum $\left\vert
0\right\rangle $ without bosons, satisfying $\hat{b}_{i\sigma }^{\dagger }%
\hat{b}_{i\sigma }=1$. This representation is equivalent to the following
operator identity between the spin and boson operators $\mathbf{\hat{S}}_{i}=%
\frac{1}{2}\mathbf{\hat{b}}_{i}^{\dagger }\mathbf{\sigma \hat{b}}_{i}$. Here
$\mathbf{\hat{b}}_{i}=(\hat{b}_{i\uparrow },\hat{b}_{i\downarrow })^{T}$ is
a bosonic spinon annihilation operator. The mean-field value is
characterized by a bosonic RVB order parameter $\Delta _{ij}^{s}=\langle
\hat{b}_{i\sigma }\hat{b}_{j,-\sigma }\rangle \neq 0$ for the
nearest-neighbor sites, which depicts the short-range AF correlation as $%
\langle \mathbf{\hat{S}}_{i}\cdot \mathbf{\hat{S}}_{j}\rangle =-\frac{1}{2}%
|\Delta _{ij}^{s}|^{2}$. At zero temperature, spinon $\mathbf{\hat{b}}$
becomes massless and Bose-condensation takes place with $\langle \mathbf{%
\hat{b}}\rangle \neq 0$,\textbf{\ }corresponding to the long-range N\'{e}el
order in $x$-$y$ plane.

To learn the property of an AF\ order with vacancies, we should generalize
the idea of \emph{spin-charge separation}, which has become a very basic
concept in understanding the doped Mott insulator related to high-$T_{c}$
cuprate\cite{anderson,KRS}. Unlike a usual quasi-particle that carries both
spin and charge quantum numbers in conventional metals, it states that the
system has two independent elementary excitations, the neutral spinon and
the spinless holon. It is assumed that usual quasi-particle excitations may
no longer be stable against the spin-charge separation mechanism once being
created, e.g., by injecting a bare vacancy into the system; and it has to
decay into more elementary spinons and holons. In other words, to
theoretically describe the introduction of a single vacancy, one has to
first annihilate a particle state $\left\vert \Psi \right\rangle $ together
with a bosonic spinon $\hat{b}_{i\sigma }$, then generate a spinless
operator $\hat{h}_{i}^{\dagger }$ denotes a holon (a vacancy) as
\begin{equation}
\hat{a}_{i\sigma }\left\vert \Psi \right\rangle =\hat{h}_{i}^{\dagger }\hat{b%
}_{i\sigma }\left\vert \Psi \right\rangle ,\ \left\vert \Psi \right\rangle
=\prod\limits_{i\sigma }\hat{b}_{i\sigma }^{\dagger }|0\rangle .
\label{wave function}
\end{equation}%
In addition, $\hat{h}_{i}^{\dagger }$ and $\hat{b}_{i\sigma }$ should
satisfy the no-double-occupancy constraint, $\hat{h}_{i}^{\dagger }\hat{h}%
_{i}+\sum_{\sigma }\hat{b}_{i\sigma }^{\dagger }\hat{b}_{i\sigma }=1$. In
contract to the case of high-$T_{c}$ cuprates, holons in bosonic $t$-$J$
model are neutral for they are actually the absence of cold atoms in a
certain site.

\begin{figure}[tbp]
\begin{center}
\includegraphics[width=0.45\textwidth]{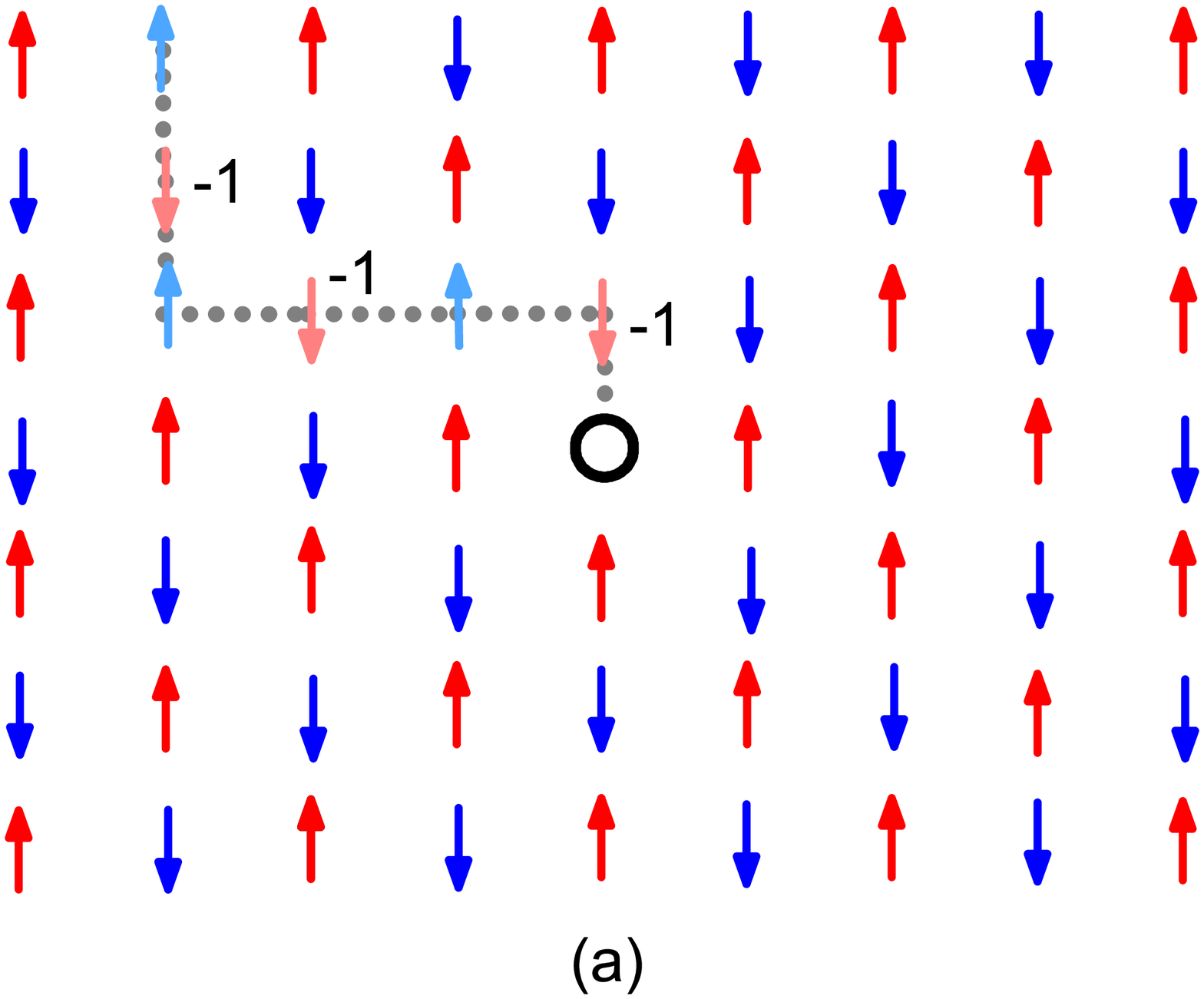} %
\includegraphics[width=0.24\textwidth]{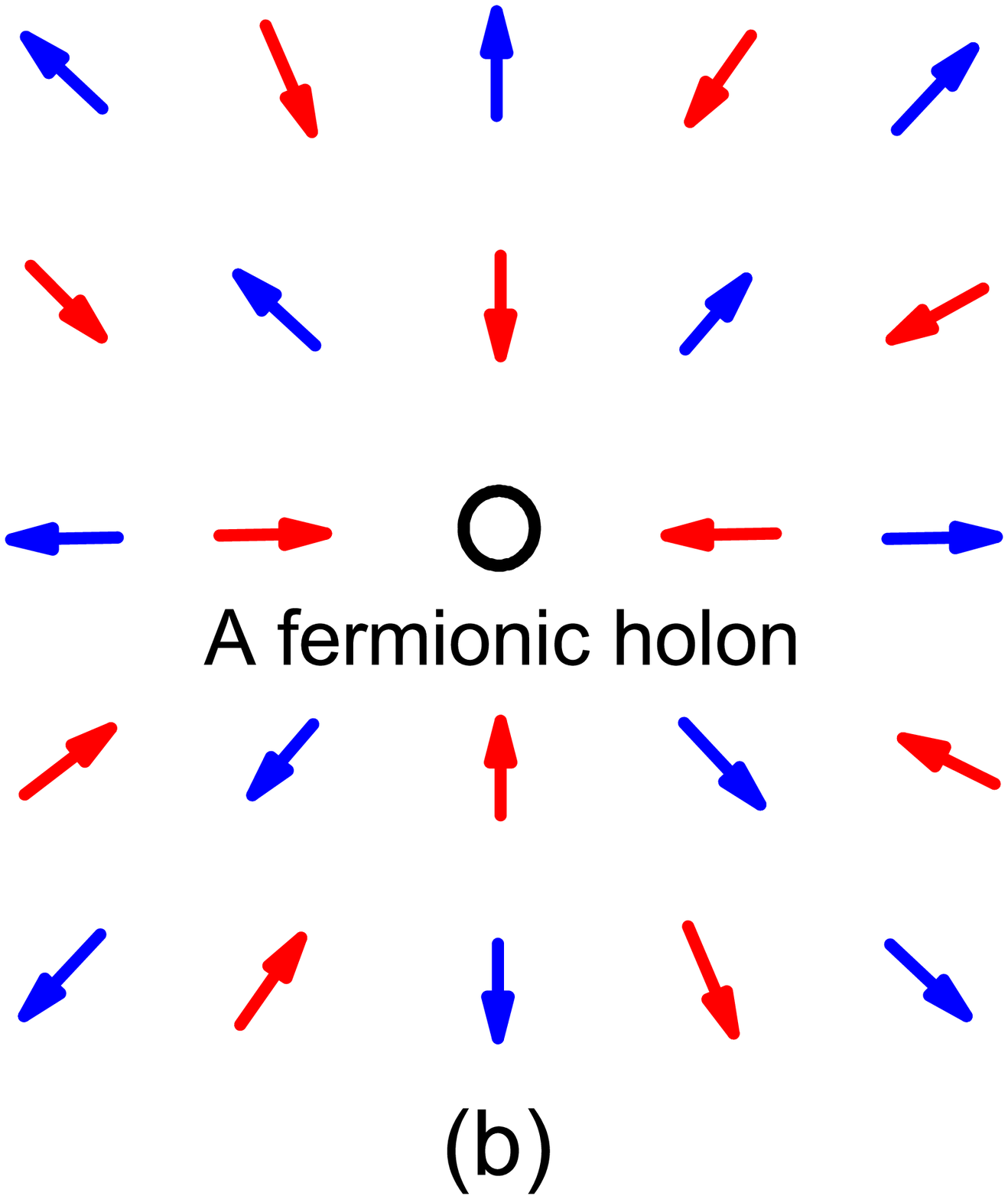} %
\includegraphics[width=0.45\textwidth]{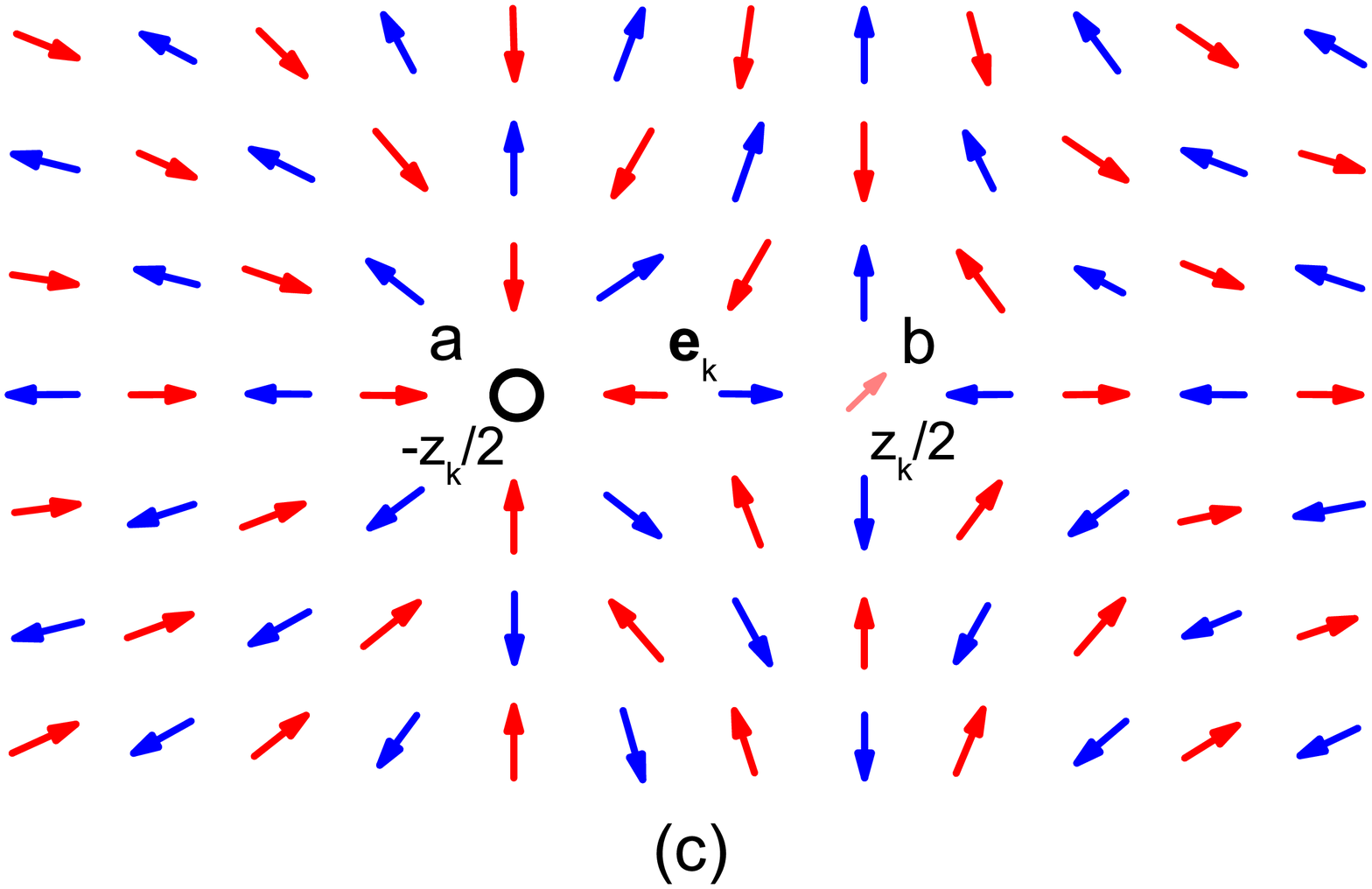}
\end{center}
\caption{(a) A schematic demonstration of the phase string effect on a
long-range AF order background. The point in $A\left( B\right) $-sublattice
is denoted by red(blue) arrows. The double-dot path is the course that a
hole (black open circle) moves through. For clarity, spins on this path is
plotted in light colors. Once the hole's hopping results a down spin on $A$%
-sublattice, there is a sign change ($-1$) over the wave function $%
\left\vert \Psi \right\rangle $, as described in the text. (b) An
illustration of a meron configuration. Due to the mutual statistics, the
vacancy has become a fermionic particle. (c) An illustration of a
holon-dipole configuration. It is a confined object composed of a
holon-meron and an anti-meron at two poles: $a\left( -z_{k}/2\right) $ and $%
b\left( z_{k}/2\right) $, connected by a branch-cut with the spatial
separation $\mathbf{\hat{e}}_{k}$ as a dipole moment. }
\label{meron}
\end{figure}

However, the situation changes when the hole moves. According to Marshall%
\cite{marshall}, the ground-state wave function of the Heisenberg
Hamiltonian for a bipartite lattice satisfies a sign rule. It requires that
flips of two antiparallel spins at nearest-neighbor sites are always
accompanied by a sign change in the wave function: $\left\vert \cdots
\uparrow \downarrow \cdots \right\rangle \mapsto \left( -1\right) \times
\left\vert \cdots \downarrow \uparrow \cdots \right\rangle $. To show the
sign effect in detail, we divide a bipartite lattice into odd ($A$) and even
($B$) sublattices and assign an extra sign $\left( -1\right) $ to every down
spin at $A$ site. When the hole initially locating at site $i$ hops onto a
nearest-neighbor site $j$, the Marshall sign rule is violated, resulting in
a string of mismatched signs on the vacancy's course\cite{trugman}. The spin
wave function is changed into%
\begin{equation*}
\left\vert \Psi \right\rangle \mapsto |\tilde{\Psi}\rangle =(-1)^{\sum_{i}%
\hat{n}_{i\in A,\downarrow }^{b}}\left\vert \Psi \right\rangle ,
\end{equation*}%
where $\hat{n}_{i\in A,\downarrow }^{b}=\hat{b}_{i\in A,\downarrow
}^{\dagger }\hat{b}_{i\in A,\downarrow }$ is the number of down spin on each
$A$-sublattice. In particular, if the hole moves through a closed path $C$
on the lattice to return to its original position, it will get a Berry phase
$(-1)^{N_{C}^{\downarrow }}$, where $N_{C}^{\downarrow }$ is the total
number of down spins \textquotedblleft encountered\textquotedblright\ by the
hole on the closed path $C$.\cite{string0,string1,string2} This process is
illustrated in Fig. (\ref{meron}a).

To deal with this unavoidable Berry phase $(-1)^{N_{C}^{\downarrow }}$ in
the ground-state wave function when there is mobile hole, we introduce a
phase-string transformation\cite{string0} $|\tilde{\Psi}\rangle \mapsto e^{i%
\hat{\Theta}}|\tilde{\Psi}\rangle =|\Psi \rangle $, where $\hat{\Theta}%
=\sum_{ij}\theta _{ij}\hat{n}_{i}^{h}\hat{n}_{j\downarrow }^{b}$, $\hat{n}%
_{i}^{h}$ and $\hat{n}_{j\downarrow }^{b}$ are occupation number operators
of the hole and down-spins, with a phase factor $\theta _{ij}=$\textrm{Im}$%
[\ln (z_{i}-z_{j})]$. Here $z\equiv x+iy$ denotes position and the
subscripts $i$ and $j$ denote lattice sites. Considering the
single-occupancy constraint, $\hat{\Theta}=-\frac{1}{2}\sum_{ij}\hat{n}%
_{i}^{h}\theta _{ij}[1-\hat{n}_{j}^{h}-\sum_{\sigma }\left( -1\right)
^{\sigma }\hat{n}_{j\sigma }^{b}]$, where $\left( -1\right) ^{\uparrow
}\equiv 1$, $\left( -1\right) ^{\downarrow }\equiv -1$. The phase-shift
factor $e^{i\hat{\Theta}}$ can also be regarded as a unitary transformation
on an arbitrary operator: $\hat{O}\mapsto e^{i\hat{\Theta}}\hat{O}e^{-i{\hat{%
\Theta}}}$.\ For example, operators of holons $\hat{h}_{i}^{\dagger }$ and
spinons $\hat{b}_{i\sigma }$\ are transformed as follow,
\begin{eqnarray*}
e^{i{\hat{\Theta}}}\hat{h}_{i}^{\dagger }e^{-i{\hat{\Theta}}} &=&\hat{h}%
_{i}^{\dagger }e^{-i\sum_{j}\theta _{ij}\hat{n}_{j}^{h}+\frac{i}{2}%
\sum_{j\sigma }\left( -1\right) ^{\sigma }\theta _{ij}\hat{n}_{j\sigma }^{b}-%
\frac{i}{2}\sum_{j}\theta _{ij}}, \\
e^{i{\hat{\Theta}}}\hat{b}_{i\sigma }e^{-i{\hat{\Theta}}} &=&\hat{b}%
_{i\sigma }e^{-\frac{i}{2}\sum_{j}\left( -1\right) ^{\sigma }\theta _{ij}%
\hat{n}_{j}^{h}}.
\end{eqnarray*}%
Now we have restored the simple spin wave function $|\Psi \rangle $ as
equation (\ref{wave function}) originally defined, but have got a set of
nontrivial\ operators, and it is just the aim of the above transformation.

Furthermore, by defining $\hat{h}_{i}^{\prime \dagger }=\hat{h}_{i}^{\dagger
}e^{-i\sum_{j}\theta _{ij}\hat{n}_{j}^{h}}$, we can see that except for a
phase factor $e^{\frac{i}{2}\sum_{j}\theta _{ij}}$, holon $\hat{h}_{i}$ and
spinon $\hat{b}_{i\sigma }$\ are symmetric,
\begin{eqnarray}
\hat{h}_{i} &\mapsto &\hat{h}_{i}^{\prime }e^{-\frac{i}{2}\sum_{j\sigma
}\left( -1\right) ^{\sigma }\theta _{ij}\hat{n}_{j\sigma }^{b}}\times e^{%
\frac{i}{2}\sum_{j}\theta _{ij}},  \notag \\
\hat{b}_{i\sigma } &\mapsto &\hat{b}_{i\sigma }e^{-\frac{i}{2}\sum_{j}\left(
-1\right) ^{\sigma }\theta _{ij}\hat{n}_{j}^{h}};  \label{t}
\end{eqnarray}%
namely, $\hat{h}_{i}\overset{h\rightarrow b,\ i\rightarrow i\sigma }{%
\longmapsto }\hat{b}_{i\sigma }$ and $\hat{b}_{i\sigma }\overset{%
b\rightarrow h,\ i\sigma \rightarrow i}{\longmapsto }\hat{h}_{i}.$ The phase
factor $e^{\frac{i}{2}\sum_{j}\theta _{ij}}$ means an additional lattice $%
\pi $-flux-per-plaquette for holons. From equation (\ref{t}), one can see
that there exists a mutual semionic statistics between holons and spinons,
where holons perceive spinons as $\pi $-vortices and vice versa. In
particular, after the transformation $\hat{h}_{i}^{\prime \dagger }=\hat{h}%
_{i}^{\dagger }e^{-i\sum_{j}\theta _{ij}\hat{n}_{j}^{h}}$, $\hat{h}%
_{i}^{\prime \dagger }$ obeys fermionic\emph{\ }anti-commutation\cite{EF},
\begin{equation*}
\{\hat{h}_{i}^{\prime \dagger },\hat{h}_{j}^{\prime }\}=\delta _{ij}.
\end{equation*}

Finally, based on the RVB ground state by considering the phase string effect%
\cite{string0,string1,string2}, the effective Hamiltonian of holons in
bosonic $t$-$J$ model can be written as%
\begin{eqnarray}
\hat{H}_{h} &=&-t_{h}\sum_{\langle ij\rangle }(e^{i\hat{a}_{ij}-i\phi
_{ij}^{0}}\hat{h}_{i}^{\prime \dagger }\hat{h}_{j}^{\prime }+H.c.)+  \notag
\\
&&\frac{\Delta }{2}\sum_{i}\left( -1\right) ^{i}\hat{h}_{i}^{\prime \dag }%
\hat{h}_{i}^{\prime }+\mu \sum_{i}\hat{h}_{i}^{\prime \dag }\hat{h}%
_{i}^{\prime },  \label{hs}
\end{eqnarray}%
where $t_{h}\approx |\Delta _{ij}^{s}|t$ is the effective hopping amplitude
of holons, $\mu $ is the chemical potential. The gauge field $\hat{a}_{ij}$
satisfy the topological constraint $\sum_{C}\hat{a}_{ij}=\pm \pi \sum_{l\in
C}(\hat{n}_{l\uparrow }^{b}-\hat{n}_{l\downarrow }^{b})$ for a closed loop $%
C $, $\phi _{ij}^{0}$ describes a $\pi $ flux per plaquette, $\sum_{\square
}\phi _{ij}^{0}=\pm \pi $. It is obvious that the difference between
fermionic $t$-$J$ model and bosonic $t$-$J$ model is the statistic of
holons: in former case, holons are bosons; and in latter case, they are
fermions.

Now we may be able to answer the question: what is the fate of a holon in an
AF order? As we have mentioned above, the AF order lying in $x$-$y$ plane
can be seen as a Bose condensation of spinons, $\langle \hat{b}_{i\sigma
}\rangle \neq 0$. According to the above effective Hamiltonian (\ref{hs}),
and the quantity $\hat{S}_{i}^{\dagger }=(-1)^{i}\hat{b}_{i\uparrow
}^{\dagger }\hat{b}_{i\downarrow }e^{i\sum_{j}\theta _{ij}\hat{n}_{j}^{h}}$,
a holon introduced here is a topological defect carrying spin\ twists and
changing its peripheral spin configuration, as a meron-like object\cite%
{meron,Berciu,Morinari1,Morinari2}, such a spin vortex (for a holon at the
origin) may be characterized by a unit vector $\mathbf{n}_{i}=\mathbf{(-}1%
\mathbf{)}^{i}\left\langle \mathbf{S}_{i}\right\rangle /S$,
\begin{equation*}
\mathbf{n}_{i}=\frac{\mathbf{r}_{i}}{\left\vert \mathbf{r}_{i}\right\vert }%
,\ \mathbf{r}_{i}^{2}=x_{i}^{2}+y_{i}^{2}.
\end{equation*}%
This meron-like spin configuration is schematically shown in Fig. (\ref%
{meron}b).

By now, one may imagine that a single hole would become a meron in a
long-range AF order. However, the answer is not quite right. For a single
meron configuration, the energy is logarithmically divergent, $E\approx
J|\Delta _{ij}^{s}|\mathrm{\ln }\left( L/a_{0}\right) $ with $L$ the size of
the system. In order to remove this infinite-energy cost, as in the present
case, each holon-meron has to \textquotedblleft nucleate\textquotedblright\
an anti-meron from the background spontaneously. Define $%
n_{i}^{x}+in_{i}^{y}=e^{i\phi _{0}+i\phi _{i}}$, with the unit vector $%
\mathbf{n}_{0}\equiv (\cos \phi _{0},\sin \phi _{0})$ as the magnetization
direction at infinity. In presence of a fermionic holon centered at $%
-z_{k}/2 $ and an anti-meron centered at $z_{k}/2$, we have\cite{kou,kou1} $%
\phi _{i}^{k}=\mathrm{Im}\ln \left[ \left( z_{i}-z_{k}/2\right) /\left(
z_{i}+z_{k}/2\right) \right] $ with $z_{k}\equiv e_{k}^{x}+ie_{k}^{y}$. By
using $\mathbf{\hat{e}}_{k}=(e_{k}^{x},e_{k}^{y})$ to denote the spatial
displacement of the holon and anti-meron centered, a \emph{dipolar} spin
configuration at a sufficiently large distance is obtained,
\begin{equation*}
\phi _{i}\approx \frac{\left( \mathbf{\hat{z}\times \hat{e}}_{k}\right)
\mathbf{\cdot r}_{i}}{\left\vert \mathbf{r}_{i}\right\vert ^{2}},\text{ }%
\left\vert \mathbf{r}_{i}\right\vert \gg \left\vert \mathbf{\hat{e}}%
_{k}\right\vert \mathbf{.}
\end{equation*}%
Thus, each pair of holons and anti-merons forms a composite, as shown in
Fig. (\ref{meron}c). In contrast to the logarithmically divergent
meron-energy, the energy a dipole $E_{d}$ becomes finite\cite{kou,kou1}, $%
E_{d}\approx J|\Delta _{ij}^{s}|\ln \left[ \left( \left\vert \mathbf{\hat{e}}%
_{k}\right\vert +a_{0}\right) /a_{0}\right] $, $\left\vert \mathbf{\hat{e}}%
_{k}\right\vert \gtrsim a_{0}$.

Physically, one may consider a bare hole (spinless holon) created by an\
annihilation operator $\hat{h}^{\dagger }$ at point $a$, then it may jump to
point $b$ via some discrete steps, being connected by a phase string in
between. Since the holon can reach $b$ through different virtual paths
originated at $a$, this singular phase string is then replaced by, or
relaxes to, a smooth \emph{dipole} configuration. An anti-meron itself is a
semi-vortex formed by condensed spinons and it is immobile. Therefore, the
hole-dipole as a whole must remain localized (self-trapped) in space. The
resulting spin configuration has a dipolar symmetry while the distortion of
the direction of magnetization is long ranged and decays as $r^{-1}$\cite%
{kou1,kou2}.

\section{Topological Fermi Liquid}

In the lightly doped region, $\delta \rightarrow 0$, there exist localized
holes that are self-trapped around anti-merons via a logarithmically
confining potential $V\left( r\right) \approx q^{2}\ln \left( \left\vert
\mathbf{r}\right\vert /a\right) $, $q^{2}=2\pi J|\Delta _{ij}^{s}|$, and
their dipolar moments are randomly distributed. Such kind of AF ordered
state with random dipoles has been studied in refs. \cite{gla,aha,che,kou}.
It is known that when $T=0$, the spin-correlation length is finite for
arbitrary doping $\delta $. This implies the destruction of long-range AF
order, as long as $\delta >0$. Consequently, AF order is limited mainly in
finite sizes, where the size $\xi $ of a domain is determined by the hole's
concentration, $\xi \approx a/\sqrt{\delta }$. Accordingly, this state has
been termed as a cluster spin glass\cite{good1,good2}. The spin glass
freezing temperature is then expected to vary as $T_{g}\varpropto \xi
^{2}\varpropto a^{2}/\delta $, below which holons tend to form a glass and
their dynamics strongly slows down.

When doping\ increases, more and more dipoles appear. As a result, the
\textquotedblleft confining\textquotedblright\ potential $V\left( r\right) $
between these dipoles will be screened by polarizations of pairs lying
between them. When this screening effect becomes so strong that even the
largest pair has to break up, holons and anti-merons are liberated to move
individually. This qualitative analysis suggests that a
localization-delocalization phase transition should occur at a critical
point $\delta _{c}$, in a fashion of the Kosterlitz-Thouless (KT)
transition. Using the renormalization group method\cite{timm,kou}, one can
determine the critical hole density $\delta _{c}=\delta _{c}(T)$ or
temperature $T_{\mathrm{de}}=T_{\mathrm{de}}(\delta )$ at which dipoles
collapse and holon-merons are \textquotedblleft
deconfined\textquotedblright\ from the bound state with anti-merons. At zero
temperature, the critical hole density has been numerically determined as
\begin{equation*}
\delta _{c}(T=0)\approx \frac{0.84}{2\pi ^{2}}=0.043.
\end{equation*}%
Therefore, a zero-temperature quantum critical point exists at $\delta
_{c}=0.043$ where hole-dipoles dissolve into holon-merons and anti-merons.

When $\delta >0.043$, fermionic holons are delocalized, so a Fermi surface
emerges in the momentum space. And since two holons always \emph{repulse}
each other, for they are merons with \emph{same} topological charges, then
these fermions cannot pair with each other and it makes the Fermi surface
stable. This is different from the case where electrons form Cooper pairs in
superconductors, because here we have topological repulsions not electronic
repulsions.

Thus, we display the global phase-diagram of bosonic $t$-$J$ model in Fig.
(3). There exists three different regions globally: region I, AF\ insulator,
$\delta =0$; region II, insulating spin-glass (SG), $\delta \in \left(
0,0.043\right) $; region III, topological Fermi-liquid (TFL) phase with
spin-liquid ground state, $\delta \in \left( 0.043,0.5\right) $. The phase
boundary that protects the physics in low doping regimes can be written as $%
k_{B}T\approx c\left( 1-2\delta \right) J,\ c$ is an integration quantity
(see Methods). Particularly, our results illustrate the emergence of an
exotic fermionic state (TFL) in a pure bosonic system.

\begin{figure}[tbp]
\begin{center}
\includegraphics[width=0.5\textwidth]{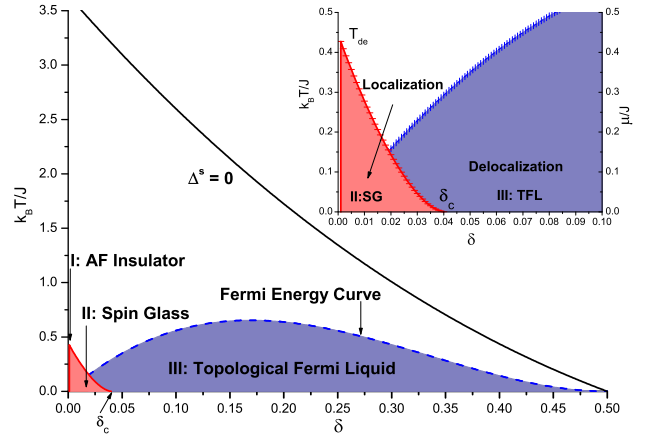}
\end{center}
\caption{The global phase diagram. The solid black line indicates
the phase boundary where $|\Delta^s_{ij}|=0$. Different regions are
demonstrated in the text. The inset depicts detailed phase
boundaries as well as the quantum
critical point $\protect\delta _{c}$ in low doping region ($\protect\delta %
<0.1$). The Fermi energy curve $\protect\mu \left( \protect\delta \right) $
implies the existence of Fermi surface of deconfined holons. It is scaled by
the superexchange energy $J$ as shown in the inset, where we have chosen $%
U/t=12$, namely, $t/J=3$ and $\Delta /J=50$.}
\label{phase_diagram}
\end{figure}

\begin{figure}[tbp]
\begin{center}
\includegraphics[width=0.5\textwidth]{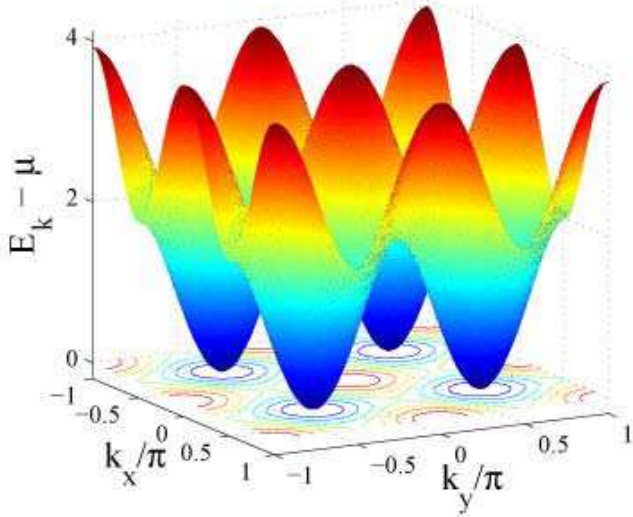}
\end{center}
\caption{The dispersion relation of the topological Fermi liquid. The
relative energy $E_{k}-\protect\mu $ of the upper branch is shown in the
unit of $t$ with $\Delta /t=20$.}
\label{dispersion}
\end{figure}

In TFL state, the effective Hamiltonian of holons can be written as
\begin{equation*}
\hat{H}_{h}=-t_{h}\sum_{\left\langle ij\right\rangle }e^{i\phi _{ij}^{0}}%
\hat{h}_{i}^{\prime \dag }\hat{h}_{j}^{\prime }+\frac{\Delta }{2}%
\sum_{i}\left( -1\right) ^{i}\hat{h}_{i}^{\prime \dag }\hat{h}_{i}^{\prime
}+\mu \sum_{i}\hat{h}_{i}^{\prime \dag }\hat{h}_{i}^{\prime },
\end{equation*}%
and the dispersion relation is
\begin{equation*}
E_{k}=\mu \pm \frac{1}{2}\sqrt{16t_{h}^{2}\left[ \cos ^{2}(k_{x})+\cos
^{2}(k_{y})\right] +\Delta ^{2}},
\end{equation*}%
as Fig. (\ref{dispersion}) shows.

\begin{figure}[tbp]
\begin{center}
\includegraphics[width=0.5\textwidth]{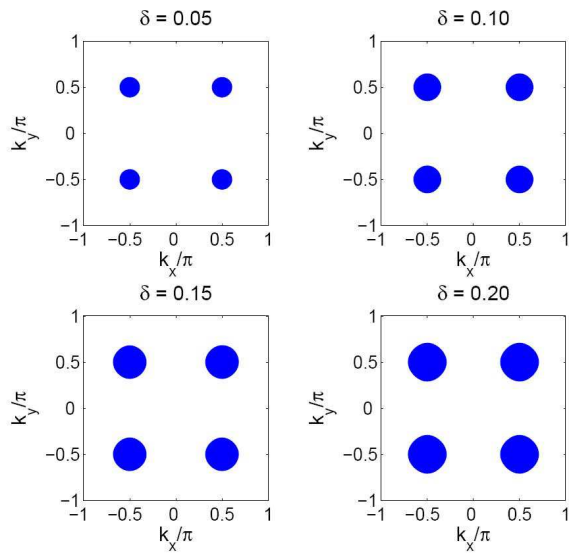}
\end{center}
\caption{Illustrations of Fermi surfaces. The doping concentrations are
shown as titles of individual plots.}
\label{Fermi surfaces}
\end{figure}

In Fig. (\ref{Fermi surfaces}), we show the evolution of Fermi surfaces by
plotting four doping cases, $\delta =0.05$, $0.1$, $0.15$ and $0.2$,
respectively. From the global phase diagram Fig. (3) and Fig. (\ref{Fermi
surfaces}), we suggest that in experiments people may observe TFL in bosonic
$t$-$J$ model of $15\%$ hole concentration, of which highest Fermi energy
with clear Fermi surfaces are predicted.

Besides, when\ holon moves, the spin configuration changes simultaneously,
thus there is no long-range AF order anymore. Instead, the magnetic
ground-state becomes a spin liquid state with invariant spin-rotation and
translation symmetry. Without anti-merons matching with holons, holons exert
an \textquotedblleft effective magnetic field\textquotedblright\ on spinons
(recall the mutual relation between the two excitations), $B_{h}=\pi \rho
_{h}$ with $\rho _{h}$ the density of holons. Hence, a new length scale is
introduced to the spinon system, which is the magnetic cyclotron length $%
l_{c}=B_{h}^{-1/2}$. $l_{c}$ will later be connected to the remaining
magnetic correlation length.

As we mentioned above, doping creates a Landau-level structure in the spinon
spectrum. Thus, low-lying spin fluctuations are expected to be sensitive to
dopants. After calculations, we find
\begin{equation*}
\chi ^{\prime \prime }\left( \mathbf{q},\omega \right) \sim \chi ^{\prime
\prime }\left( \mathbf{Q}_{0},\omega \right) e^{{-\left\vert \mathbf{q}-%
\mathbf{Q}_{0}\right\vert ^{2}l}_{c}^{2}/2}.
\end{equation*}%
Here $\chi ^{\prime \prime }\left( \mathbf{q},\omega \right) $ is defined as
($\beta =1/k_{B}T$)
\begin{equation}
\chi ^{\prime \prime }\left( \mathbf{q},\omega \right) =\frac{1}{2}\left(
1-e^{-\omega \beta }\right) \int dtd\mathbf{r}e^{i\left( \omega t-\mathbf{%
q\cdot r}\right) }\left\langle \mathbf{S}\left( \mathbf{r},t\right) \cdot
\mathbf{S}\left( 0,0\right) \right\rangle  \label{spin correlation}
\end{equation}%
where $\left\langle \mathbf{S}\left( \mathbf{r},t\right) \cdot \mathbf{S}%
\left( 0,0\right) \right\rangle $ is the spin-spin correlation function.
This is a Gaussian type around the AF wave-vector $\mathbf{Q}_{0}=(\pm \pi
,\pm \pi )/a_{0}$ with a width $l_{c}^{-1}$. Consequently, it determines the
spin-spin correlation in the real space as $\cos \left( \mathbf{Q}_{0}\cdot
\mathbf{r}\right) e^{-\left\vert \mathbf{r}\right\vert ^{2}/\xi ^{2}}$,
where the correlation length $\xi =\sqrt{2}l_{c}=a_{0}\sqrt{2/\left( \pi
\delta \right) }$ is in the same order of the average hole-hole distance\cite%
{weng}. Namely, doped holes break up the long-range AF correlation into
short-range AF fragments within a length scale of $\xi $.

In contrast, the ground-state of fermionic $t$-$J$ model is a
superconducting state of condensed holons, in which spin liquid state is
always masked by holons' condensation\cite{Lee}. While in bosonic $t$-$J$
model, it is the topological Fermi liquid state with massless fermionic
excitations, in which one can easily detect properties of this spin liquid
state. For these reasons, the observation of this exotic quantum state (TFL
with short range spin correlation) in bosonic $t$-$J$ model may pave an
alternative approach to verify the microscopic theory of high $T_{c}$
superconductivity.

\section{Detections and Summary}

In region I and II, the system is an insulator in the first place, so the
momentum distribution is universal and isotropic in the Brillouin zone
(without considering the trapping and boundary conditions). This feature can
be detected by measuring density profile measurement\cite{profile}, or via
Bragg spectroscopy\cite{Bragg} to observe the momentum spectrum.
Additionally, in the N\'{e}el state (I), the condensate exhibits long range
AF order, while in the spin glass state (II), this long ranged magnetic
order disappears. This difference can be distinguished through spatial
spin-spin correlations, $\langle \hat{S}_{z}\left( \mathbf{r}_{1}\right)
\hat{S}_{z}\left( \mathbf{r}_{2}\right) \rangle $. As demonstrated in refs.%
\cite{noise1,noise2}, using a probe laser beam which goes through the
condensate, one can measure the phase shift or change of polarization of the
outgoing beam $\langle \hat{X}_{\text{out}}\rangle $ to obtain the
magnetization $\langle \hat{X}_{\text{out}}\rangle \propto \langle \hat{M}%
_{z}\rangle \propto \int d\mathbf{r}\phi \left( \mathbf{r}\right) \langle
\hat{S}_{z}\left( \mathbf{r}\right) \rangle $, where $\phi \left( \mathbf{r}%
\right) $ is the spatial intensity profile of the laser beam. The quantum
noise $\langle \hat{X}_{\text{out}}^{2}\rangle \propto \langle \hat{M}%
_{z}^{2}\rangle \propto \int d\mathbf{r}_{1}d\mathbf{r}_{2}\phi \left(
\mathbf{r}_{1}\right) \phi \left( \mathbf{r}_{2}\right) \langle \hat{S}%
_{z}\left( \mathbf{r}_{1}\right) \rangle \langle \hat{S}_{z}\left( \mathbf{r}%
_{2}\right) \rangle $ reveals the atomic correlations in the system. This
method can directly examine the existence of 2D AF correlations.

In region III, we have demonstrated that one natural characteristic of TFL
is the existence of Fermi surfaces in the Brillouin zone. Esslinger et al.%
\cite{Fermi surface} had successfully observed the Fermi surface in a 3D
optical lattice filled with fermionic atoms. Similarly, one may also observe
these Fermi levels in our bosonic system, as shown in Fig. (\ref{Fermi
surfaces}). Besides, the short-range AF fluctuations may also be observed.
As equation (\ref{spin correlation}) indicates, the spin dynamic structure
factor as well as the dynamic spin susceptibility function are linked with
spin-spin correlations. The latter one reflects the effective short-range
magnetic correlation length $\xi $.

To sum up, in this paper, we first demonstrate the implementation of bosonic
$t$-$J$ model in optical superlattices filled with a two-component BEC. And
for the first time, we systematically discussed the possible quantum phases
of this model upon doping. A key feature is the mutual semionic statistics
between the two elementary excitations of the system, holons and spinons.
When there are only a few holes around, they tend to form hole-dipoles,
locally changing the underlying spin texture into meron-anti-meron pairs. At
low doping, the deformed spin configuration loses the long-range order, and
there emerges a spin glass state. When holes are prevailing, through a
quantum phase transition ($\delta _{c}=0.043$)\ that frees those
self-localized holons in spin glass state, a strange topological Fermi
liquid with Fermi surface appears in a purely bosonic system, as a
significant result of the mutual statistics between holons and spinons. We
also list accessible experimental approaches to verify and detect these
novel theoretical predictions.

\section{Methods}

\subsection{Perturbation Theory in Large $U$ Limit}

The two-component Bose-Hubbard model in a biased superlattice can be
generally written as%
\begin{eqnarray}
\hat{H} &=&-t\sum_{\left\langle ij\right\rangle \sigma }[\hat{a}_{i\sigma
}^{\dag }\hat{a}_{j\sigma }+H.c.+\frac{\Delta }{2t_{\sigma }}\left( \hat{n}%
_{i\sigma }-\hat{n}_{j\sigma }\right) ]+  \notag \\
&&U\sum_{i}\hat{n}_{i\uparrow }\hat{n}_{i\downarrow }+\frac{U^{\prime }}{2}%
\sum_{i}\hat{n}_{i\sigma }\left( \hat{n}_{i\sigma }-1\right) ,  \label{BH1}
\end{eqnarray}%
where $t=\sqrt{16/\pi }E_{r}\left( V/E_{r}\right) ^{3/4}e^{-2\sqrt{V/E_{r}}}$%
\cite{Duan} with atomic recoil energy $E_{r}=\hbar ^{2}k^{2}/2m$, $k$ is the
wave vector of laser, $m$ is the atomic mass, $V$ describes external
potential. The inter-species repulsion is defined as $U=\sqrt{8/\pi }%
ka_{s}E_{r}\left( V/E_{r}\right) ^{3/4}$ with $a_{s}$ the $s$-wave
scattering length among species, and the intra-species repulsion is $%
U^{\prime }=\sqrt{8/\pi }ka_{s}^{\prime }E_{r}\left( V/E_{r}\right) ^{3/4}$,
$a_{s}^{\prime }$ is the corresponding scattering length. This term vanishes
at exact unit-filling, but can contribute in higher-order tunneling
processes.

In the large $U$ limit ($U\gg t$) and nearly unit-filling $n\lesssim 1$,
which is actually a prerequisite for building the aforementioned
superlattices and is naturally satisfied, the hopping term can be treated as
a perturbation. Thus, we introduce two projection operator: $\mathcal{P}$
projects the initial Hilbert space onto the single-occupancy subspace ($%
\left\vert 0\right\rangle _{i},\left\vert \uparrow \right\rangle
_{i},\left\vert \downarrow \right\rangle _{i}$), $\mathcal{Q}=1-\mathcal{P}$
projects onto the double-occupancy subspace ($\left\vert \uparrow \downarrow
\right\rangle _{i}$). We further divide equation (\ref{BH1}) into $\hat{H}%
_{0}+\hat{T}_{\text{mix}}$, where $\hat{H}_{0}$ describes processes in $%
\mathcal{P}$ subspace and $\hat{T}_{\text{mix}}$ mixes the upper and lower
band via virtual tunnelings.

Applying canonical transformation $\hat{H}_{\text{eff}}=e^{-\hat{S}}\hat{H}%
e^{\hat{S}}$ and eliminating first order terms of $\hat{T}_{\text{mix}}$, we
find
\begin{eqnarray*}
\hat{H}_{\text{eff}} &=&\hat{H}_{0}+\frac{1}{2}\left[ \hat{T}_{\text{mix}},%
\hat{S}\right] , \\
\hat{S} &=&\sum_{mn}\left\vert m\right\rangle \left\langle m\right\vert
\left( \frac{\hat{T}_{\text{mix}}}{E_{n}-E_{m}}\right) \left\vert
n\right\rangle \left\langle n\right\vert .
\end{eqnarray*}%
The projector $\left\vert \epsilon \right\rangle \left\langle \epsilon
\right\vert $ can be chosen as $\mathcal{P}$ and $\mathcal{Q}$. Energy
differences between upper and lower band are $E_{\mathcal{Q}}-E_{\mathcal{P}%
}\approx U\pm \Delta ,\frac{U^{\prime }}{2}\pm \Delta $ thus $\hat{S}=(%
\tilde{U}^{-1}+2\tilde{U}^{\prime -1})(\mathcal{P}\hat{T}\mathcal{Q}-%
\mathcal{Q}\hat{T}\mathcal{P})$ with $\tilde{U}=U-\Delta ^{2}/U$, $\tilde{U}%
^{\prime }=U^{\prime }-\Delta ^{2}/U^{\prime }$. After calculating the
detailed terms, we can obtain the effective Hamiltonian as
\begin{eqnarray}
\hat{H}_{\mathrm{eff}} &=&-t\sum_{\left\langle ij\right\rangle \sigma }(\hat{%
a}_{i\sigma }^{\dag }\hat{a}_{j\sigma }+H.c.)+  \notag \\
&&\sum_{\left\langle ij\right\rangle }[J_{z}\hat{S}_{i}^{z}\hat{S}%
_{j}^{z}-J_{\perp }(\hat{S}_{i}^{x}\hat{S}_{j}^{x}+\hat{S}_{i}^{y}\hat{S}%
_{j}^{y})],  \label{Heff1}
\end{eqnarray}%
where
\begin{equation*}
J_{z}=\frac{4t^{2}}{\tilde{U}}-\frac{8t^{2}}{\tilde{U}^{\prime }},J_{\perp }=%
\frac{4t^{2}}{\tilde{U}}.
\end{equation*}%
When $a_{s}$ and $a_{s}^{\prime }$ do not differ very much (this condition
can somehow be realized in experiments),
\begin{equation*}
J_{z}\approx -\frac{4t^{2}}{\tilde{U}}+\epsilon
\end{equation*}%
with $\epsilon $ denoting a small spatial inhomogeneity. In fact, this small
inhomogeneity is necessary in our following discussions, for it can be seen
as an analogy for the effective inter-layer-coupling in high-$T_{c}$
cuprates.

At the first glance, it seems that bosons (say, $\left\vert \uparrow
\right\rangle $) in shallow sites may hop to neighboring deep sites more
easily due to lower\ energy barrier they feel, comparing their counterparts (%
$\left\vert \downarrow \right\rangle $) in the reverse process. However,
thanks to spin-dependent controlling techniques, which can be achieved by
adjusting parameters of incident light beams, one can regulate tunneling
amplitudes of different species, making atoms in different sites feel a
similar hopping matrix element, $t_{\uparrow }=t_{\downarrow }\equiv t$.
This is the reason we straightforwardly set an identical hopping term $t$ at
the beginning.

Meanwhile, to achieve an effective AF superexchange interaction, we need to
set $\Delta =\sqrt{2}U$ ($\tilde{U}=-U$) to change signs of coefficients: $%
J_{z}\approx 4t^{2}/U+\epsilon $, $J_{\perp }=-4t^{2}/U$ in equation (\ref%
{Heff1}). As a result, the global coefficient of spin interaction becomes
positive, $J\approx 4t^{2}/U$, as equation (\ref{bosonic tJ}) asks.

Under these circumstances, we finally derive the bosonic $t$-$J$ model with
a small inhomogeneity in $z$ direction, which can be formally written as
equation (\ref{bosonic tJ}). Quite recently, two papers \cite{Barthel,DW}
suggested a similar proposal and made detailed discussions.

\subsection{Mean field calculation from slave-particle approach}

This representation is equivalent to the following operator identity between
the spin and boson operators. Defining $\hat{a}_{i\sigma }=\hat{h}_{i}^{\dag
}e^{i\hat{\Theta}_{i\sigma }^{\text{string}}}\hat{b}_{i\sigma }$ and $%
\mathbf{\hat{S}}_{i}=\frac{1}{2}\mathbf{\hat{b}}_{i}^{\dagger }\mathbf{%
\sigma \hat{b}}_{i}$, where $\hat{h}_{i}^{\dag }$ creates a fermionic holon,
leaving a non-local phase string and $\hat{b}_{i\sigma }$ annihilates a
bosonic (Schwinger-boson) spinon at site $i$, we rewrite equation (\ref%
{bosonic tJ}) as $\hat{H}=\hat{H}_{h}+\hat{H}_{s}$:
\begin{eqnarray*}
\hat{H}_{h} &=&-t\sum_{\left\langle ij\right\rangle \sigma }e^{i\hat{A}%
_{ij}^{f}}\hat{h}_{i}^{\dag }\hat{h}_{j}e^{i\hat{A}_{ji}^{f}}\hat{b}%
_{j\sigma }^{\dag }\hat{b}_{i\sigma }, \\
\hat{H}_{s} &=&-\frac{J}{2}\sum_{\left\langle ij\right\rangle \sigma \sigma
^{\prime }}e^{i\hat{A}_{ij}^{h}}\hat{b}_{i\sigma }^{\dag }\hat{b}_{j,-\sigma
}^{\dag }e^{i\hat{A}_{ji}^{h}}\hat{b}_{j,-\sigma ^{\prime }}\hat{b}_{i\sigma
^{\prime }},
\end{eqnarray*}%
Here $\hat{\Theta}_{i\sigma }^{\text{string}}=\frac{1}{2}(\hat{\Phi}%
_{i}^{b}-\sigma \hat{\Phi}_{i}^{h})$ is a topological phase with the
contribution from spinon number $\hat{n}_{\sigma }^{b}$, $\hat{\Phi}%
_{i}^{b}=\sum_{l\neq i,\sigma }\left( -1\right) ^{\sigma }\mathrm{Im}\left[
\ln \left( z_{i}-z_{l}\right) \right] \hat{n}_{l\sigma }^{b}$ and the
holon's ($\hat{n}^{h}$) contribution, $\hat{\Phi}_{i}^{h}=\sum_{l\neq i}%
\mathrm{Im}\left[ \ln \left( z_{i}-z_{l}\right) \right] \hat{n}^{h}.$ And $%
\hat{A}_{ij}^{f}$ and $\hat{A}_{ij}^{h}$ describe quantized fluxes bounded
to spinons and holons. We then introduce a mean-field bosonic RVB order
parameter $\Delta ^{s}=\sum_{\sigma }\langle e^{-i\sigma \hat{A}_{ij}^{h}}%
\hat{b}_{i\sigma }\hat{b}_{j,-\sigma }\rangle $, and take the constraint on
total spinon number and dilute effect into account, we apply mean-field
approximation on $\hat{H}$ and solve two self-consistent equations:
\begin{eqnarray*}
\delta  &=&2-\frac{1}{N}\sum_{k}\lambda _{k}E_{k}^{-1}\coth \left( \beta
E_{k}/2\right) , \\
\Delta ^{s} &=&\frac{1-2\delta }{N}\sum_{k}\xi
_{k}^{2}J_{s}^{-1}E_{k}^{-1}\coth \left( \beta E_{k}/2\right) ,
\end{eqnarray*}%
to obtain the phase boundary
\begin{equation*}
k_{B}T=2J\left( 1-2\delta \right) \left( 2-\delta \right) \left( \ln \frac{%
3-\delta }{1-\delta }\right) ^{-1}
\end{equation*}%
that separates $\Delta ^{s}=0$ and $\Delta ^{s}\neq 0$. Here $N$ is the
total lattice number. Detailed definitions can be found in ref. \cite%
{string1}.

\begin{acknowledgments}
The authors thank Z.-Y. Weng, Y.-B. Zhang, B. Wu, W.-M. Liu for helpful
discussions and comments. This research is supported by NCET and NFSC Grant
no. 10574014, 10874017.
\end{acknowledgments}


\begin{thebibliography}{99}
\bibitem{EF} The emergent fermionic excitations has been pointed out in
another full bosonic system in Ref. \cite{ef}. However, the mechanism in
Ref. \cite{ef} is much different from ours, both the Marshall sign rule and
the phase string effect.
\end{thebibliography}
\end{document}